\documentclass[aps,prl,onecolumn,preprintnumbers,superscriptaddress,nofootinbib,floatfix,10pt]{revtex4}

\usepackage{amsfonts,amssymb,stmaryrd,latexsym,amsmath,braket}
\usepackage{graphicx,subfigure,blindtext}
\usepackage{bm}
\usepackage{bookmark}
\usepackage{times}
\usepackage{slashed}
\DeclareMathOperator{\sgn}{sgn}

\begin{document}

\title{Time fractals and discrete scale invariance with trapped ions}

\author{Dean~Lee}
\affiliation{Facility for Rare Isotope Beams and Department of Physics and
Astronomy, Michigan State University, MI 48824, USA}

\author{Jacob~Watkins}
\affiliation{Facility for Rare Isotope Beams and Department of Physics and
Astronomy, Michigan State University, MI 48824, USA}

\author{Dillon~Frame}
\affiliation{Facility for Rare Isotope Beams and Department of Physics and
Astronomy, Michigan State University, MI 48824, USA}

\author{Gabriel~Given}
\affiliation{Facility for Rare Isotope Beams and Department of Physics and
Astronomy, Michigan State University, MI 48824, USA}

\author{Rongzheng~He}
\affiliation{Facility for Rare Isotope Beams and Department of Physics and
Astronomy, Michigan State University, MI 48824, USA}

\author{Ning~Li}
\affiliation{Facility for Rare Isotope Beams and Department of Physics and
Astronomy, Michigan State University, MI 48824, USA}

\author{Bing-Nan~Lu}
\affiliation{Facility for Rare Isotope Beams and Department of Physics and
Astronomy, Michigan State University, MI 48824, USA}

\author{Avik~Sarkar}
\affiliation{Facility for Rare Isotope Beams and Department of Physics and
Astronomy, Michigan State University, MI 48824, USA}


\begin{abstract}
We show that a one-dimensional chain of trapped ions can be engineered
to produce a quantum mechanical system with discrete scale invariance and
fractal-like time dependence.    By discrete scale invariance we mean a system
that replicates itself under a rescaling of distance for some scale factor, and a time fractal is a signal that is invariant under the rescaling
of time.
These features are reminiscent  of the Efimov effect, which has been predicted
and observed in bound states of three-body systems. We demonstrate that discrete
scale invariance in the trapped ion system can be controlled with two independently
tunable parameters.  We also discuss the extension to $n$-body states where  the
discrete scaling symmetry has an exotic heterogeneous structure.   The results we present
can be realized using
currently available technologies developed for trapped ion quantum systems. 
\end{abstract}

\maketitle

In this work we show how to construct a one-dimensional system of trapped
ions with discrete scale invariance and fractal-like time
dependence.
In classical systems scale invariance arises when the scale transformation
acting on spatial coordinates, $r \rightarrow \lambda r,$ is a symmetry of
the dynamics.  This arises naturally if the Hamiltonian transforms homogeneously
under rescaling.  When the Hamiltonian is quantized, however, this scale
invariance cannot persist for bound state solutions with discrete energy
levels.  Instead, the scale invariance is broken through a quantum scale
anomaly.  An analogous effect occurs in relativistic field theories and is
responsible for the mass gap in the spectrum of non-Abelian gauge theories
such as quantum chromodynamics. 

While the quantum scale anomaly spoils invariance
under a general scale transformation, it may preserve the symmetry associated
with a discrete set of scale transformations.  This was first described by
Efimov for the bound state spectrum of three bosons with short-range
interactions tuned to infinite scattering length \cite{Efimov:1971zz,Efimov:1993a,Bedaque:1998kg,Bedaque:1998km}.
 See also Ref.~\cite{Coon:2002sua} for a review
of anomalies in quantum mechanics and the attractive $1/r^2$ potential.
 Efimov trimers were first observed experimentally through the loss rate
of trapped ultracold cesium atoms \cite{Kraemer:2006Nat}, and a more direct
observation has been made using the Coulomb explosion of helium trimers \cite{Kunitski:2015qth}.
As the underlying physics is of universal character, the application and
generalization of the Efimov effect has been considered in various settings,
including nuclear physics \cite{Bedaque:1999ve,Hagen:2013jqa}, bound states
with more than three particles \cite{Platter:2004pra,Hammer:2006ct,vonStecher:2009a,vonStecher:2011zz,Carlson:2017txq},
systems with reduced dimensions \cite{Nishida:2011ew,Moroz:2013kf,Happ:2019}, quantum
magnets \cite{Nishida:2012hf}, molecules with spatially-varying
interactions \cite{Nishida:2012by}, and Dirac fermions in graphene \cite{Ovdat:2017lho}.

We demonstrate that quantum scale anomalies can be produced with trapped
ion quantum systems.  We start with a one-dimensional chain of ions in
a radio-frequency
trap with qubits represented by two hyperfine ``clock''
states.  Such systems have been investigated by the trapped ion group at the University of Maryland using   $^{171}$Yb$^{+}$ ions  \cite{Zhang:2017a,Zhang:2017b}.
  Similar efforts have been pioneered by trapped ion groups at ETH Z{\"u}rich, Freiburg, Innsbruck,  Mainz, Stockholm, and the Weizmann Institute. 
Off-resonant laser beams are used
to drive stimulated Raman transitions for all ions in the trap.  This induces
effective interactions between
all qubits with
a power-law dependence on separation distance.
 We define the vacuum state as the state  with $\sigma^z_i=1$ for all $i$.
 We use interactions of the form $\sigma^x_{i}\sigma^x_{j}+\sigma^y_{i}\sigma^y_{j}$,
to achieve the hopping of spin excitations.  We then use a
$\sigma^z_{i}\sigma^z_{j}$ interaction to produce  a two-body potential
felt by pairs of spin excitations, and we also consider an external one-body
potential coupled to $\sigma^z_{i}$. 

We can view each spin
excitation  with $\sigma^z_i=-1$ as a bosonic particle
at site $i$ with hardcore interactions preventing multiple occupancy.   In
this language, the Hamiltonian we consider  has the form 
\begin{align}
H =\frac{1}{2} \sum_{i}\sum_{j \ne i} J_{ij}[b^{\dagger}_i b_j +b^{\dagger}_j
b_i] & + \frac{1}{2}\sum_{i}\sum_{j \ne i}V^{}_{ij}b^{\dagger}_i b_i b^{\dagger}_j
b_j \nonumber\\
& + \sum_{i}U^{}_{i}b^{\dagger}_i b_i+C,
\end{align}
where $b_i$ and $b^{\dagger}_i$ are annihilation and creation operators for
the hardcore bosons on site $i$.  See the Supplemental Materials for a derivation of this Hamiltonian.  The parameter $C$ is just an overall energy constant.  The  hopping coefficients $J_{ij}$ have
the asymptotic form
$J_{ij} = J_0/|r_i-r_j|^{\alpha}$, where $r_i$ is the position of qubit $i$.
 For the purposes of this study, we assume $J_{ij}$ to have exactly this
form for $i \ne j.$ 
 Similarly, the two-body potential coefficients $V_{ij}$ have the
asymptotic form $V_{ij} = V_0/|r_i-r_j|^{\beta}$.  In this work we assume
$V_{ij}$ to have exactly this form for $i \ne j$.  We consider the case where
the lattice of ions is uniform and large, and we start with a constant potential
$U_i$ chosen so that bosons with zero momentum have zero energy.
Both positive (anti-ferromagnetic) and negative (ferromagnetic) values can
be realized for $J_0$ and $V_0$. The exponents $\alpha$ and $\beta$ can in
principle vary in the range between $0$ and $3$. However, in practice the
range between $0.5$ and $1.8$ is favored in order to enhance coherence times
and reduce experimental drifts \cite{Zhang:2017b}.  

We now add to $U_i$ a deep attractive potential at some chosen site $i_0$
that traps and immobilizes one boson at that site.  Without loss of generality,
we take the position of that site to be the origin and add a constant to
the Hamiltonian so that the energy of the
trapped boson is zero.  We then consider the dynamics
of a second boson that feels the interactions with this fixed boson at the
origin.  In order to produce a Hamiltonian with classical
scale invariance, we choose $\beta=\alpha-1$.  Then at low energies, our
low-energy Hamiltonian for the second boson has the form
\begin{equation}
H(p,r)=2J_0\sin(\alpha\pi/2)\Gamma(1-\alpha)|p|^{\alpha-1}+
 \frac{V_0}{|r|^{\alpha-1}},
\end{equation} 
where we omit corrections of size $O(p^{2})$. We are interested in the case
where both $J_0$ and $V_0$ are negative.   In that case we find an infinite
tower of even parity and odd parity bound states.  We label the bound state
energies as $E^{(n)}_{+}$ and  $E^{(n)}_{-}$, respectively, for nonnegative
integers $n$.  As expected, our quantized system has a quantum scale anomaly
and we are left with two discrete scale symmetries, $r \rightarrow \lambda_{+}
r$ for even parity and  $r \rightarrow \lambda_{-} r$ for odd
parity. Correspondingly, the bound state energies follow a simple geometrical
progression, $E^{(n)}_{+}=E^{(0)}_+ \lambda_{+}^{-n}$ and $E^{(n)}_{-}=E^{(0)}_-
\lambda_{-}^{-n}$.  In the Supplemental Materials we provide details of the
discrete scale invariance for general $\alpha$. For the special case $\alpha=2$,
the scale factors are  $\lambda_{\pm} = \exp (\pi/\delta_{\pm}),$ where \begin{align}
\delta_+ = \frac{V_0}{J_0\pi} \coth (\delta_+ \pi/2), \; \; 
\delta_- = \frac{V_0}{J_0\pi} \tanh (\delta_- \pi/2).
\end{align}   

\begin{figure}
\centering
\includegraphics[width=8.5cm]{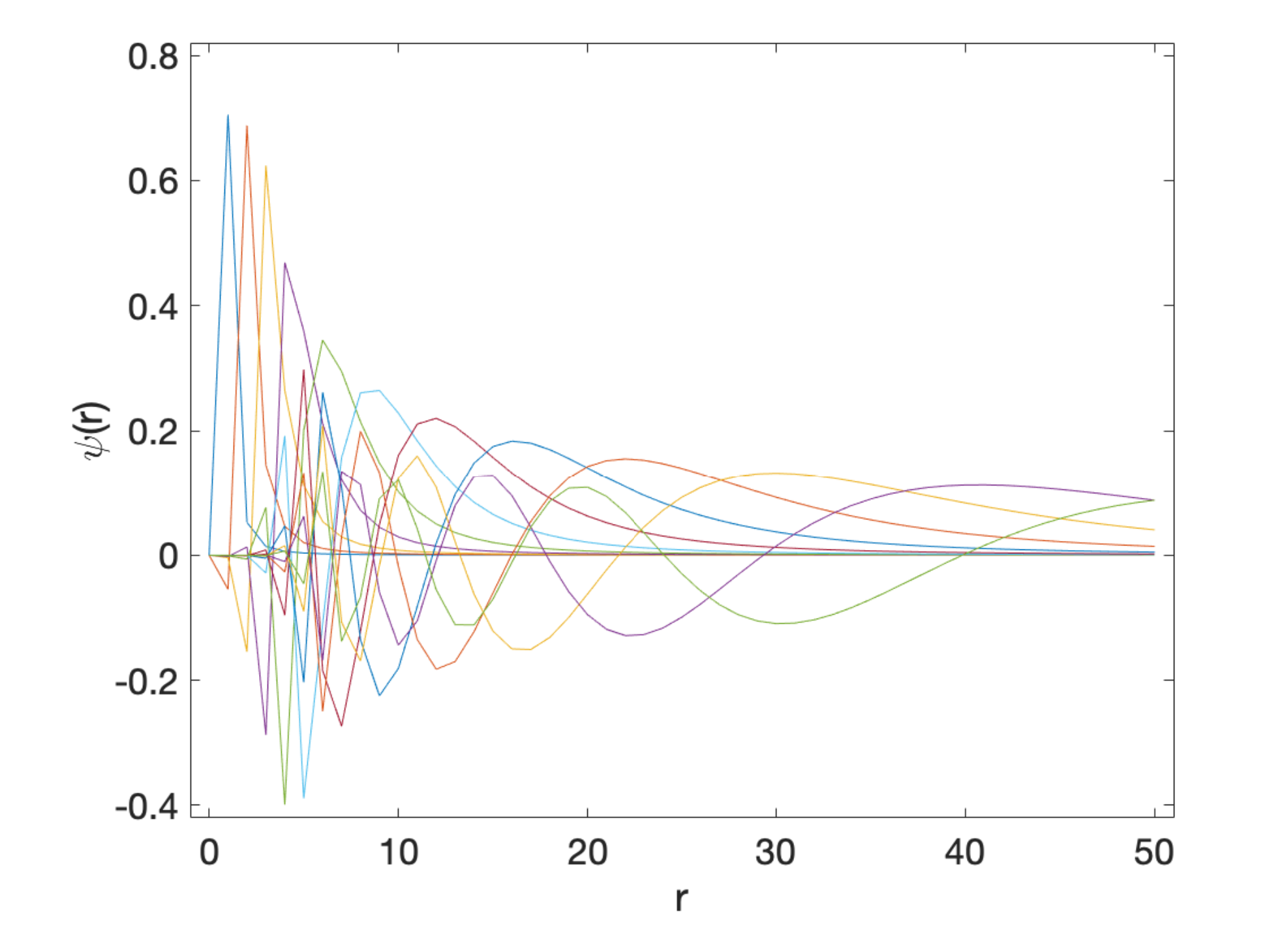}
\caption{{\bf Bound state wave functions.} Plot of the normalized wave functions
for the first twelve even-parity bound states for the case $\alpha = 2$,
$\beta =  1$, $J_0 = -1$,
and $V_0 = -30$.  We plot the region $r>0$.  All quantities are in dimensionless lattice units.}
\label{even_wavefunctions}
\end{figure}
 
In contrast with most other systems with a quantum scale anomaly, we note
that the properties of our ion trap system can be tuned using two different
adjustable parameters, $V_0/J_0$ and $\alpha$.  This is convenient for probing
a wide range of different phenomena exhibiting discrete scaling symmetry.
 In the following we will work in lattice units where physical quantities are multiplied by powers
of the lattice spacing to make the combination  dimensionless and have set
$\hbar = 1$.  As an example, consider a system with $\alpha = 2$, $\beta = 1$, $J_0 =
-1$, and $V_0
= -30$.  The wave functions for the first twelve even-parity bound states
are shown in Fig.~\ref{even_wavefunctions}. 
We plot
the normalized wave function for $r>0$.
 We see clear evidence of discrete scale invariance emerging as we approach
zero energy. In Table~\ref{energies} we show the energies for the first fourteen
even-parity and odd-parity bound states and the ratios between consecutive
energies. For comparison, at the bottom we show the predictions for these
ratios as we approach zero energy at infinite volume.  We see that the agreement
is quite good.

\begin{table*}
\begin{tabular}{ c c c c c}
\hline 
$n$ & $E^{(n)}_+$ & $E^{(n-1)}_+/E^{(n)}_+$ & $E^{(n)}_-$ & $E^{(n-1)}_-/E^{(n)}_+$
\tabularnewline
\hline 
$0$ & $-27.05304149$ & $-$ & $-26.5188669$ & $-$ \\
$1$ & $-11.93067205$ & $2.267520336$ & $-11.79861873$ & $2.247624701$ \\
$2$ & $-6.977774689$ & $1.709810446$ & $-6.919891389$ & $1.705029468$ \\
$3$ & $-4.553270276$ & $1.5324754$ & $-4.521425357$ & $1.530466798$ \\
$4$ & $-3.139972298$ & $1.450098869$ & $-3.120231851$ & $1.449067112$ \\
$5$ & $-2.233327278$ & $1.405961557$ & $-2.220194049$ & $1.405386998$ \\
$6$ & $-1.617052389$ & $1.381110033$ & $-1.607920414$ & $1.380786033$ \\
$7$ & $-1.182654461$ & $1.367307563$ & $-1.176124883$ & $1.367134084$ \\
$8$ & $-0.869406941$ & $1.360300229$ & $-0.864656962$ & $1.360221377$ \\
$9$ & $-0.640405903$ & $1.357587332$ & $-0.636916042$ & $1.357568195$ \\
$10$ & $-0.471738446$ & $1.357544438$ & $-0.469161911$ & $1.357561276$ \\
$11$ & $-0.347112043$ & $1.359037968$ & $-0.345207121$ & $1.359073675$ \\
$12$ & $-0.254996818$ & $1.361240684$ & $-0.253589633$ & $1.361282464$ \\
$13$ & $-0.187011843$ & $1.363532996$ & $-0.18597462$ & $1.363571189$ \\
\hline
theory & -- & $\lambda_{+}=1.3895595319$ &  -- & $\lambda_{-}=1.3895595319$
\\
\hline
\end{tabular}
\caption{{\bf Bound state energies.} Energies for the first fourteen even-parity
and odd-parity bound states and ratios between consecutive energies for the
case $\alpha = 2$,
$\beta = 1$, $J_0 = -1$,
and $V_0 = -30$.    For comparison we show the theoretical predictions for
the ratios $\lambda_{+}$ and $\lambda_{-}$ as we approach zero energy at
infinite volume.}
\label{energies}
\end{table*}
One intriguing question is how discrete scale invariance could persist in
quantum many-body systems.  It has been demonstrated numerically that the
Efimov effect extends beyond bosonic trimers and describes the properties
of $n$-boson systems with the same discrete scaling factor \cite{Platter:2004pra,Hammer:2006ct,vonStecher:2009a,vonStecher:2011zz,Carlson:2017txq}
.  As we will see, something quite different happens in the trapped ion system.
 Let us start from a particular bound state of the two-body system and ask
what happens when we introduce a third boson that is weakly bound and very
far from the origin. The effective Hamiltonian for the third boson  
contains a potential energy that is doubled due to interactions of the weakly-bound
third boson with the two other bosons.
As a result of the stronger attractive interaction, the geometric scaling
factors $\lambda_{\pm}$ for the third boson will be smaller than for the
two-body system.
This argument can be generalized to describe weakly-bound states for the
general $n$-body system.  The effective potential for the $n^{\rm th}$ boson
will be a factor of $n-1$ times larger, and thus the scaling of the $n$-body
energies relative to each $(n-1)$-body threshold is different from the scaling
of the $k$-body bound states for each $k$ between $1$ and $n$.  The properties
of
these exotic systems with heterogeneous discrete invariance will be investigated
further in future work.

Let us now consider an initial state $|S\rangle = \sum_{n=0}^{N-1} |\psi^{(n)}_+\rangle,$
where we sum over the first $N$ even-parity two-boson bound states $|\psi^{(n)}_+\rangle$
 with equal weight.  We choose the even-parity states, but we could just
as easily choose odd-parity states.  The phase convention for each $|\psi^{(n)}_+\rangle$
is chosen
so that the tail of the wave function is real and positive at large $r$.
We note that the time dependent amplitude $A(t)={\rm Re} [\langle S | \exp(-iHt)
| S \rangle]$ is invariant under the rescaling  $t \rightarrow \lambda^{\alpha-1}_{+}
t$, thus endowing it with the properties of a time fractal. \ The time fractal
is particularly interesting for the case when $\lambda^{\alpha-1}_{+}$ is
an integer so that each of the higher frequencies in $A(t)$ are integer multiples
of the lower frequencies.

\begin{figure*}
\centering
\includegraphics[width=8cm]{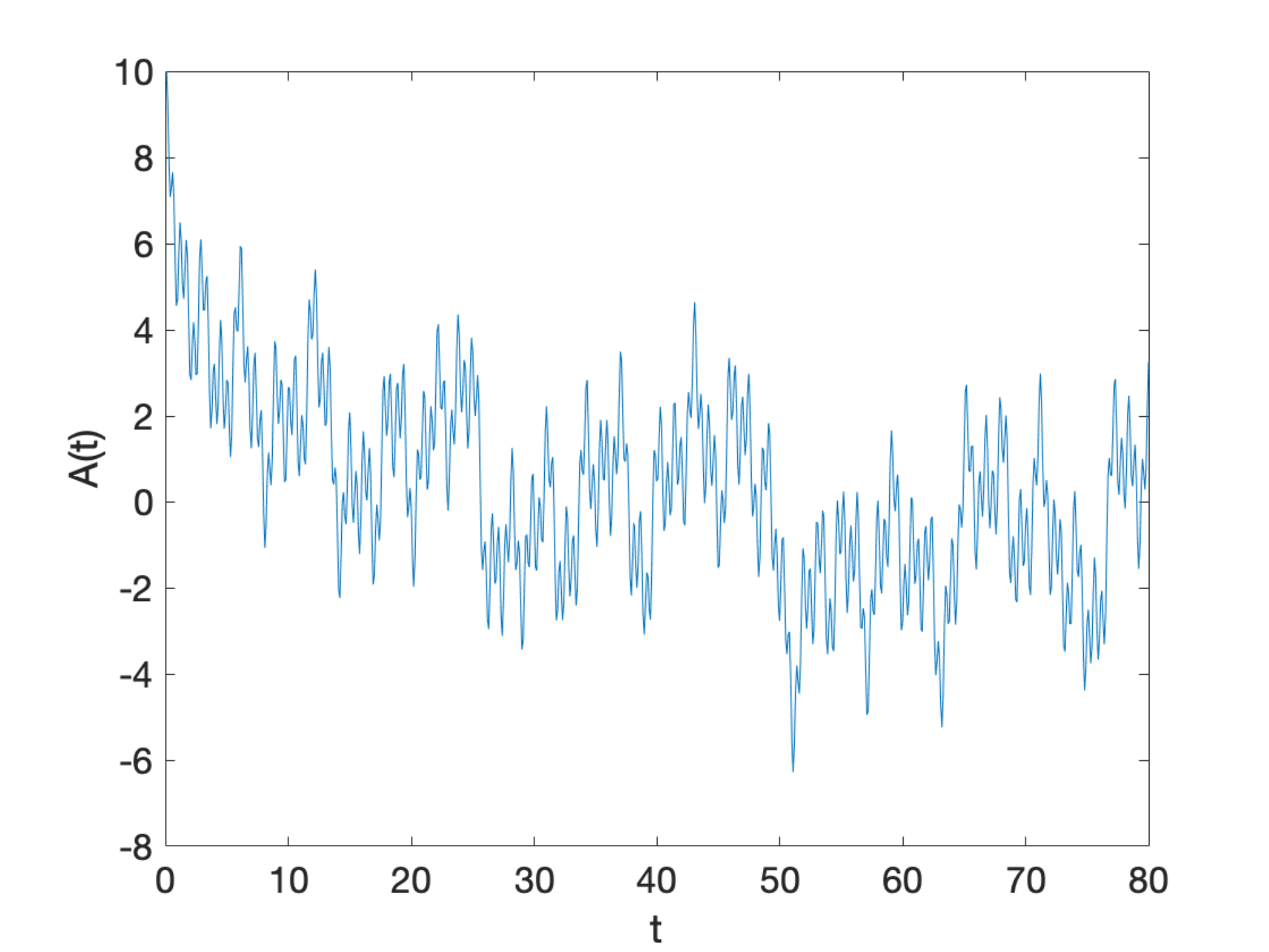}
\includegraphics[width=8cm]{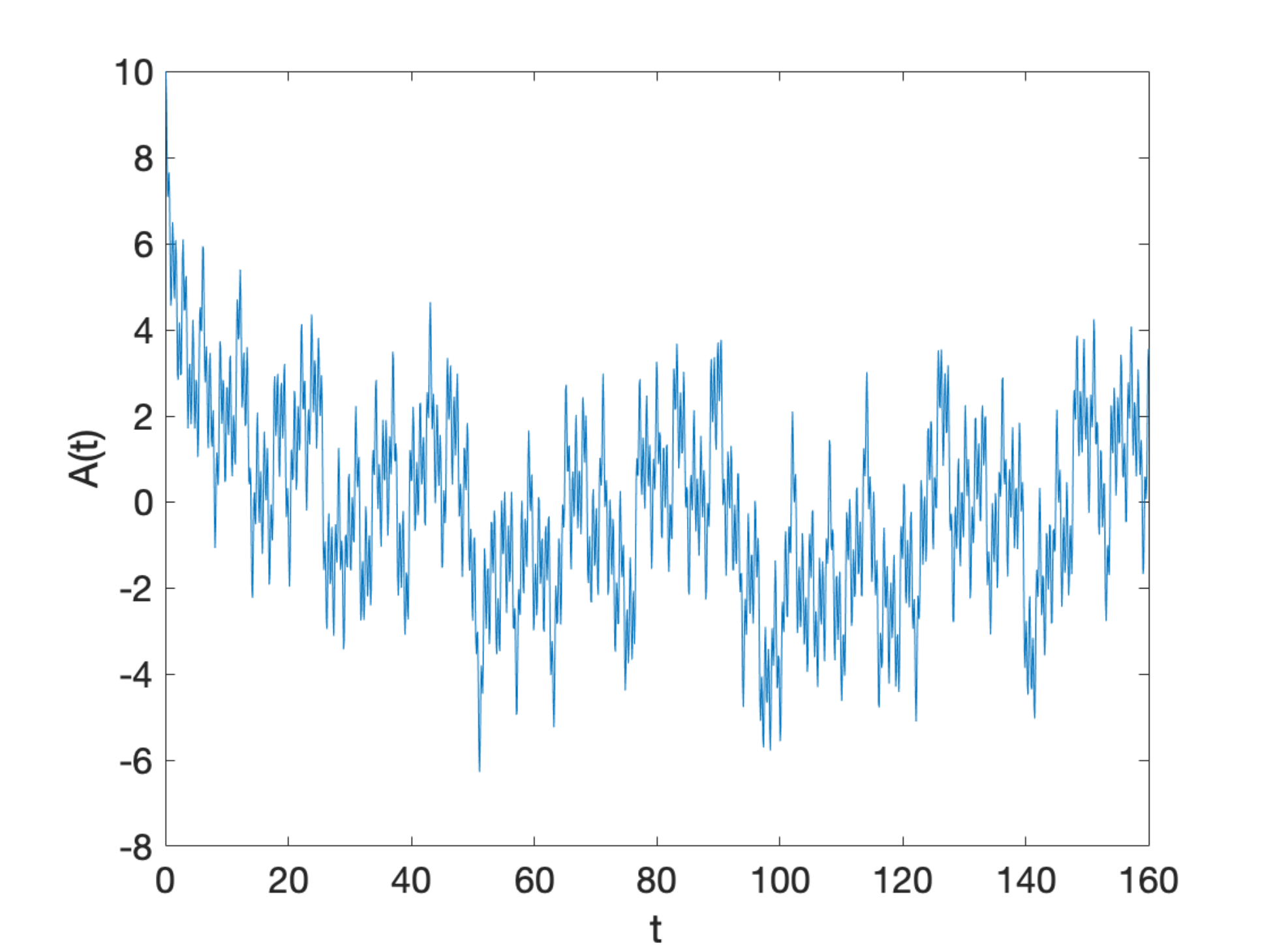}
\includegraphics[width=8cm]{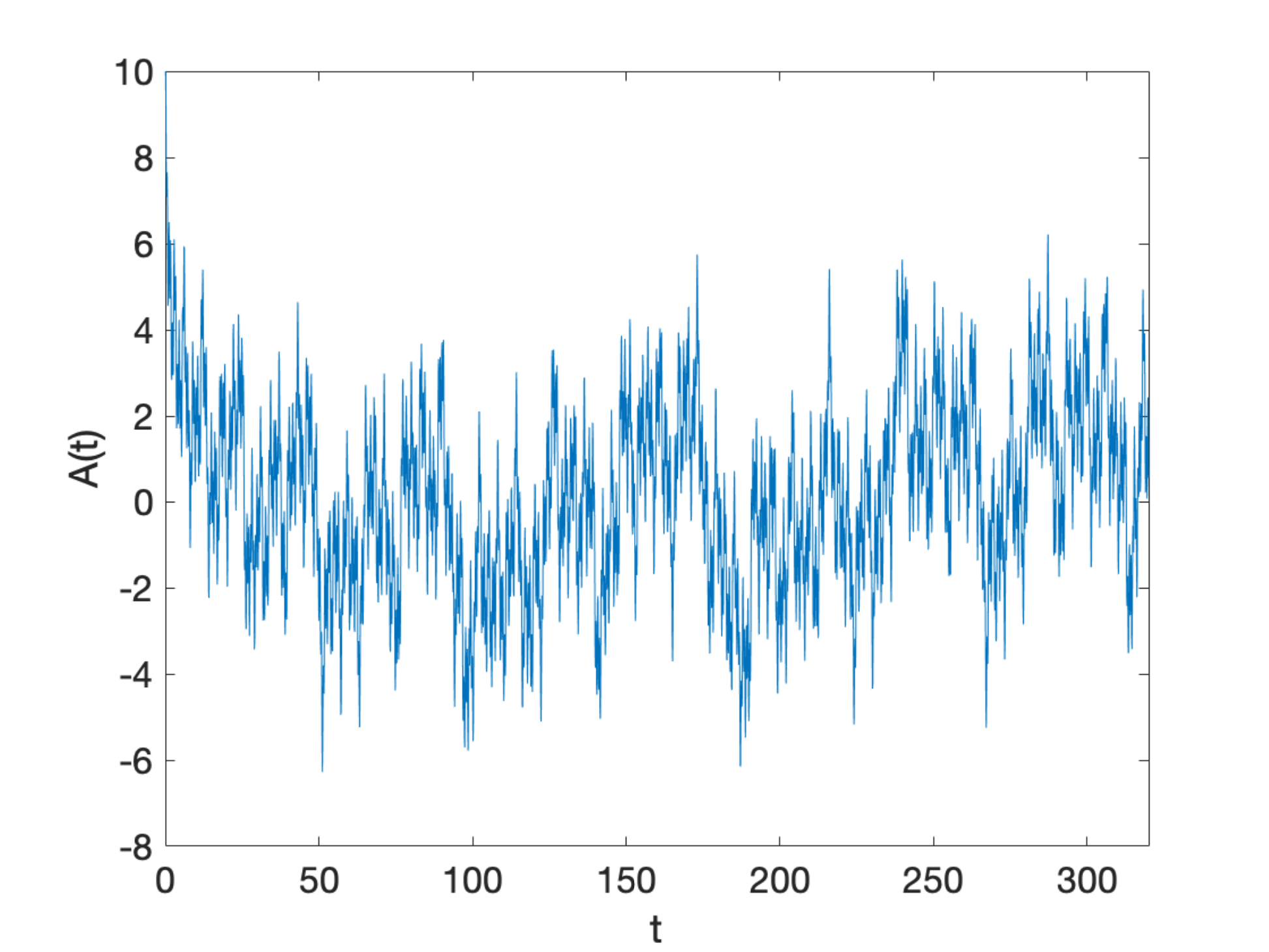}
\includegraphics[width=8cm]{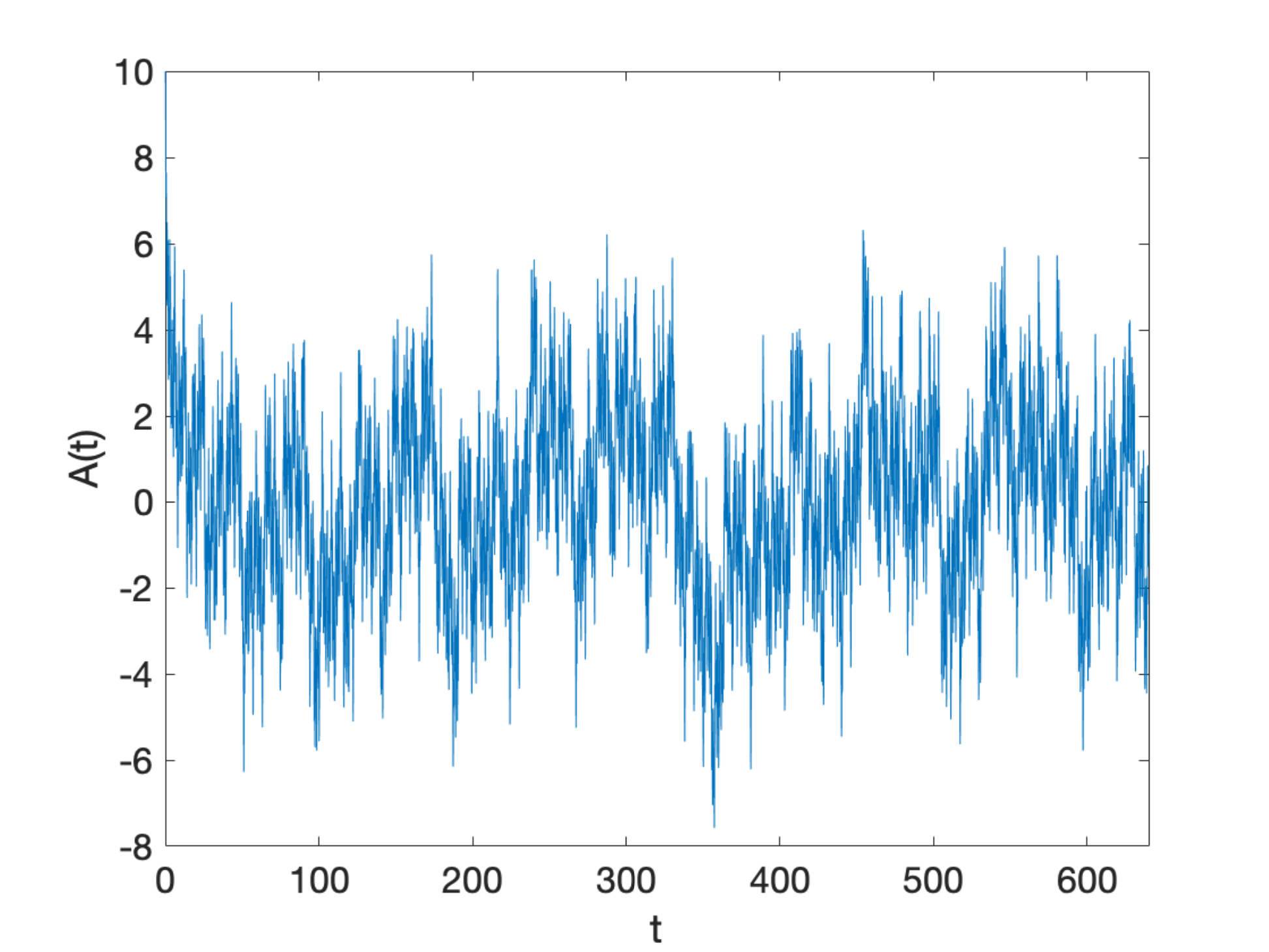}
\caption{{\bf Time fractals.} The amplitude $A(t)$ is displayed over the
range  from $t=0$ to $80$ in the upper left, $t=0$ to $160$ in the upper
right, $t=0$ to $320$ in the lower left, and $t=0$ to $640$ in the lower
right. All quantities are in dimensionless lattice units.}
\label{fractal}

\end{figure*}

For the case $\alpha = 2$ and $J_0=-1$, we can produce the time scaling
factor $\lambda^{\alpha-1}_+=\lambda_+=2$ by setting $V_0=-14.2388293$. 
In Fig.~\ref{fractal}
we show the amplitude $A(t)$ ranging  from $t=0$ to $80$ in the upper left,
$t=0$ to $160$ in the upper right, $t=0$ to $320$ in the lower left, and
$t=0$ to $640$ in the lower
right.  Aside from small deviations, we see that the time dependence shows
fractal-like self-similarity when we zoom in or out by a scale factor
very close to $2$.  The best fit for the scale factor is approximately $1.9$.
In the Supplemental
Materials we show how a time fractal can be realized experimentally using
quantum interference on a trapped ion quantum system.  

The time fractals that we have discussed are closely related to the Weierstrass
function
$w(x)=\sum_{n=0}^\infty a^n \cos(b^n  \pi x)$.  Weierstrass showed that this function
is continuous everywhere but differentiable nowhere when $0<a<1$, $b$ is
an odd integer, and $ab > 1 + 3\pi/2$~\cite{Weierstrass:1886}. Hardy extended
the proof to any $0<a<1<b $ and $ab\ge 1$ ~\cite{Hardy:1916}.  We note that
$aw(bx)$ equals $w(x)$   plus the smooth function $\cos(\pi x)$, and this
suggests that the fractal dimension of the Weierstrass function should given
by \cite{Hunt:1998}
\begin{equation}
D = 2 + \frac{\log a}{\log b}.
\end{equation}  
This result for the fractal dimension is confirmed by the box-counting method
for determining fractal dimensions~\cite{Kaplan:1984}. 
 
Our initial state $|S\rangle = \sum_{n=0}^{N-1} |\psi^{(n)}_+\rangle$
produces the fractal-like amplitude
\begin{equation}
A(t) =\sum_{n=0}^{N-1}\cos (E^{(n)}_+t) =\sum_{n=0}^{N-1}\cos (\epsilon_+ \lambda_{+}^{-n}t).
\end{equation}
In the limit of large $N$, our choice of parameters corresponds to the limiting
case $a \rightarrow 1$ and $b = \lambda_{+}^{}$, with $x =\epsilon_+\lambda_+^{-N+1}t/{\pi}$.
 Therefore, the fractal dimension for our time fractal will be $D=2$. If
we instead choose the initial state to have the form
$|S(a)\rangle = \sum_{n=0}^{N-1} a^{n/2}|\psi^{(n)}_+\rangle $  
for $a < 1,$ then in the limit $N \rightarrow \infty$, the fractal dimension
will be
\begin{equation}
D = 2 + \frac{\log a}{\log \lambda_{+}}.
\end{equation}  

There are many interesting related phenomena that one can explore in connection
with time fractals and the dynamics of systems with discrete scale invariance.
 One fascinating topic is the adiabatic evolution of a system with discrete
invariance
as the interactions are varied slowly. Another is the response of a system
with discrete scale
invariance when driven in resonance with one of its bound state
energies. In this letter we have shown that  the intrinsic power-law interactions
of the trapped ion system make it an ideal system for exploring the physics
of quantum scale anomalies, discrete scale
invariance, and time fractals.  There are clearly many directions that one
can explore in this new area, and we look forward to working with others
to develop further applications and experimental realizations of many of
these concepts.     

{\it We are grateful for discussions with Zohreh Davoudi, Chao Gao, Pavel
Lougovski,
Titus Morris, Thomas Papenbrock, and Raphael Pooser.  We acknowledge financial
support from the U.S. Department of Energy (DE-SC0018638 and DE-AC52-06NA25396).
 Computational resources were provided by the
Julich Supercomputing Centre at Forschungszentrum J\"ulich, Oak Ridge
Leadership Computing Facility, RWTH Aachen, and Michigan State University.}

\newpage
\renewcommand{\thefigure}{S\arabic{figure}}
\setcounter{figure}{0}

\section{Supplemental Material}

\subsection*{Trapped ion Hamiltonian}
\begin{figure}[!ht]
\centering

\includegraphics[width=6cm]{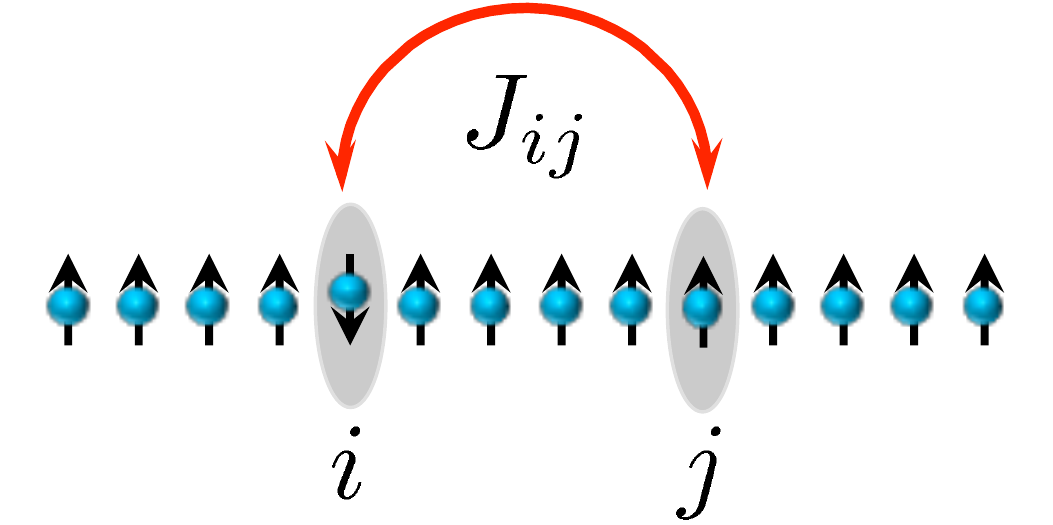}
\caption{{\bf Hopping coefficient $J_{ij}$.} This sketch shows the action
of the hopping coefficient $J_{ij}$ for a single particle between sites $i$
and $j$.  Particle excitations correspond with sites where $\sigma^z=-1$.}
\label{J_ij}
\end{figure}

\begin{figure}[!ht]
\centering
\includegraphics[width=6cm]{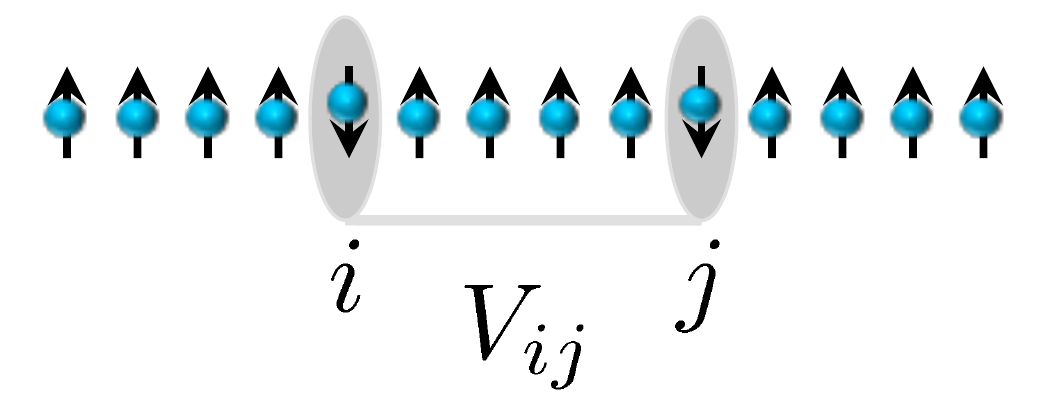}
\caption{{\bf Interaction potential $V_{ij}$.} This sketch shows the two-body
interaction potential $V_{ij}$ between two particles at sites $i$ and $j$.
Particle excitations correspond with sites where $\sigma^z=-1.$}
\label{V_ij}
\end{figure}

For our one-dimensional trapped ion system, the Hamiltonian we consider is
 \begin{equation} 
 H = T + V_2+U+C, 
 \end{equation} 
where 
\begin{align}
 T & = \frac{1}{4}\sum_{i}\sum_{j \ne i} J_{ij}(\sigma^x_{i} \sigma^x_{j}+\sigma^y_{i}
\sigma^y_{j}), \\
 V_2 & = \frac{1}{8}\sum_{i}\sum_{j \ne i}V_{ij}(1-\sigma^z_{i})( 1-\sigma^z_{j}
), \\
 U & =\frac{1}{2}\sum_{i}U_{i}( 1-\sigma^z_{i}),
\end{align}
and $C$ is a constant. 
We regard each spin
configuration  with $\sigma^z=-1$ as a particle excitation.  Thus $T$ corresponds
to the hopping of a single particle, $V_2$ is a two-particle interaction,
and $U$ is a one-particle potential. In Fig.~\ref{J_ij} we show a sketch
of the action
of the hopping coefficient $J_{ij}$.  In Fig.~\ref{V_ij} we show a sketch
of the two-body interaction potential $V_{ij}$.  Without loss of generality,
we assume
that both $J_{ij}$ and $V_{ij}$ are symmetric in the indices $i,j.$

We can reorganize the $\sigma^z$ terms as 
\begin{equation}
V_2 +U=\frac{1}{8}\sum_{i}\sum_{j \ne i}V_{ij}\sigma^z_{i}\sigma^z_{j}
-\frac{1}{2}\sum_{i}U'_i\sigma^z_{i}+C',
\end{equation}
where 
\begin{equation}
U'_i = U_i+\frac{1}{2}\sum_{i \ne j}V_{ij},
\end{equation}
and 
\begin{equation}
C'=\frac{1}{8}\sum_{i}\sum_{j \ne i}V_{ij} +\frac{1}{2}\sum_{i}U_{i}.
\end{equation}
We can view each spin
excitation  with $\sigma^z_i=-1$ as a bosonic particle
at site $i$ with hardcore interactions preventing multiple occupancy.  When
expressed in terms of hardcore boson annihilation and creation operators,
the Hamiltonian becomes 
\begin{equation}
H =\frac{1}{2} \sum_{i}\sum_{j \ne i} J_{ij}[b^{\dagger}_i b_j+b^{\dagger}_j
b_i] + \frac{1}{2}\sum_{i}\sum_{j \ne i}V^{}_{ij}b^{\dagger}_i b_i b^{\dagger}_j
b_j+ \sum_{i}U^{}_{i}b^{\dagger}_i b_i+C.
\label{Hamiltonian}
\end{equation}
\subsection*{Dispersion relation}
\begin{figure}[!ht]
\centering
\includegraphics[height=5cm]{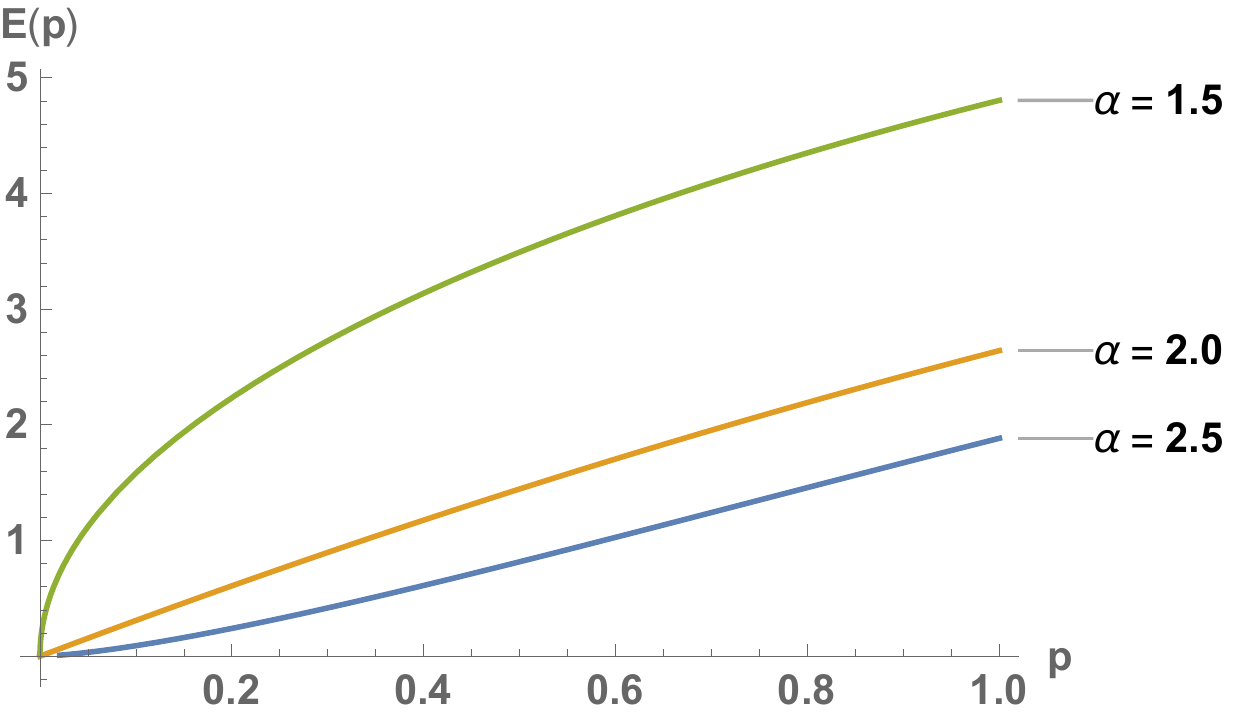}
\caption{{\bf Dispersion relation.} Plot of the dispersion relation $E(p)$
versus $p$ for $J_0=-1$. }
\label{dispersion_relation}
\end{figure}

We assume that the ions lie on a one-dimensional lattice with uniform spacing.
 There will be some distortion at the edges of the trap, but since our interest
is in bound states with some degree of spatial localization, these edge effects
can be minimized by placing the system at the middle of a trap with many
ions.  We
work in lattice units where physical quantities are multiplied by powers
of the lattice spacing to make the combination  dimensionless and have also
set $\hbar = 1$.  We start
with the case where the potential $U_i$ is set to equal
\begin{equation}
U_{i} =- \sum_{j \ne i} J_{ij}=- \sum_{j \ne i} \frac{J_0}{|r_i-r_j|^{\alpha}}.
\end{equation}  By computing the expectation value of the Hamiltonian for
a single boson with momentum $p$, we find that the energy of a single boson
with momentum $p$ is 
\begin{equation}
E(p) = 2J_0\sum_{n>0} \frac{\cos(pn)-1}{n^\alpha}  =J_0 \left[ {\rm Li}_{\alpha}(e^{ip})+{\rm
Li}_{\alpha}(e^{-ip})-{\rm 2Li}_{\alpha}(1) \right],
\end{equation}
where ${\rm Li}_{\alpha}$ is the polylogarithm function of order $\alpha$.
 We find that for $\alpha<3$,
\begin{equation}
E(p)= 2J_0\sin(\alpha\pi/2)\Gamma(1-\alpha)|p|^{\alpha-1}+J_0\zeta(\alpha-2)p^{2}+O(p^4),
\label{dispersion}\end{equation}
where $\zeta$ is the Riemann zeta function. We note that the special case
$\alpha=2$ corresponds with a linear dispersion
relation, which has important theoretical connections to relativistic fermions
as well as electrons in graphene.  In Fig.~\ref{dispersion_relation} we plot
the dispersion relation $E(p)$
versus $p$ for $J_0=-1$ and $\alpha=1.5,2.0,2.5$.
 
\subsection*{Two-body system}
We now introduce a single-site deep trapping potential with large coefficient
$u>0$ at some site $i_0$ that traps and immobilizes one hardcore boson at
that site,   
\begin{equation}
U_{i} = - \sum_{j \ne i} \frac{J_0}{|r_i-r_j|^{\alpha}}-u\delta_{i,i_0}.
\end{equation}
We also subtract a constant from the Hamiltonian so that the energy of the
trapped boson is exactly zero. We then consider the dynamics
of a second boson that feels the interactions with this fixed boson at $i_0$.
 In order to simplify our notation, let the position of the fixed boson be
$r_{i_0}=0$.  Let us now set $\beta=\alpha-1$.  Then at low energies, our
low-energy Hamiltonian for the second boson has the form
\begin{equation}
H(p,r)=2J_0\sin(\alpha\pi/2)\Gamma(1-\alpha)|p|^{\alpha-1}+
 \frac{V_0}{|r|^{\alpha-1}},
\end{equation} 
with corrections of size $O(p^{2})$. This follows from the result for $E(p)$
in Eq.~(\ref{dispersion}) and that the interaction between the particles
has the form $V_{ij} = V_0/|r_i-r_j|^{\beta} = V_0/|r_i-r_j|^{\alpha-1}$.
We note that this Hamiltonian has classical
scale invariance.  

We define by analytic continuation the Fourier transforms
of the functions $|r|^\eta$, 
\begin{equation}
\frac{1}{\sqrt{2\pi}}\int_{-\infty}^\infty dr e^{ipr} |r|^\eta=-\sqrt{\frac{2}{\pi}}|p|^{-1-\eta}\Gamma(1+\eta)\sin(\eta\pi/2).
\end{equation}
We also compute the Fourier transforms of
$\sgn(r)|r|^\eta$, where $\sgn(r)$ is the sign function,\begin{equation}
\frac{1}{\sqrt{2\pi}}\int_{-\infty}^\infty dr e^{ipr} \sgn(r)|r|^\eta=i\sqrt{\frac{2}{\pi}}\sgn(p)|p|^{-1-\eta}\Gamma(1+\eta)\cos(\eta\pi/2).
\end{equation}
In the zero energy limit, the quantum Hamiltonian $H(p,r)$ exhibits a renormalization-group
limit cycle with even-parity $(+)$ and odd-parity $(-)$ wave functions at
zero energy,
\begin{align}
\psi_{\rm +}(r)  & = \frac{1}{2}\left(|r|^{i\delta_+} + |r|^{-i\bar{\delta}_+}\right),\\
 \psi_{\rm -}(r) &  =  \frac{1}{2}\sgn(r)\left(|r|^{i\delta_{-}} + |r|^{-i\bar{\delta}_{-}}\right),
\end{align}
where $\delta_{\pm}$ are solutions to the constraints, 
\begin{align}
2J_0\delta_+\Gamma(1-\alpha)\sin(\alpha\pi/2)\Gamma(i\delta_+)\sinh(\delta_+\pi/2)&
=V_{0}
\Gamma(2-\alpha+i\delta_+)\cos((\alpha-i\delta_+)\pi/2), \\
2J_0\delta_-\Gamma(1-\alpha)\sin(\alpha\pi/2)\Gamma(i\delta_-)\cosh(\delta_-\pi/2)&
=iV_{0}
\Gamma(2-\alpha+i\delta_-)\sin((\alpha-i\delta_-)\pi/2).
\end{align}
For the particular case $\alpha =2$, this constraint simplifies to   
\begin{align}
\delta_+ = \frac{V_0}{J_0\pi} \coth (\delta_+ \pi/2), \; \; 
\delta_- = \frac{V_0}{J_0\pi} \tanh (\delta_- \pi/2).
\end{align}  These solutions for the case $\alpha =2$ are real whenever $V_0/J_0$
is positive. When
$V_0/J_0 \gg \pi$,
these  are both very well approximated by
\begin{equation}
\delta_+ \approx \delta_- \approx   \frac{V_0}{J_0\pi}.
\end{equation}

In the cases where $\delta_+$ and $\delta_-$ are real, the discrete scale
invariance of the renormalization-group limit cycle can
be seen by writing
\begin{align}
\psi_{\rm {+}}(r)  = \cos[\delta_{+} \ln(|r|)], \;  \; \;\psi_{\rm {-}}(r)
= \sgn(r)\cos[\delta_{-} \ln(|r|)].
\end{align} 
Under the scale transformations $r \rightarrow \lambda_{\pm} r,$ we have
\begin{align}
\psi_{\rm {+}}(r) & \rightarrow   \cos[\delta_{+} \ln(|r|)+\delta_{+} \ln(\lambda_{+})],
\\
\psi_{\rm {-}}(r) & \rightarrow    \sgn(r)\cos[\delta_{-} \ln(|r|)+\delta_{-}
\ln(\lambda_{-})].
\end{align}
The wave functions remain invariant up to an overall minus sign if
we let\begin{align}
\lambda_{+} = \exp (\pi/\delta_{+}), \; \;
\lambda_{-} = \exp (\pi/\delta_{-}).
\end{align}
The bound state energies also respect this discrete scale symmetry.  Under
the scale transformation $r \rightarrow \lambda_{\pm} r$, the energy scales
as $E_{\pm} \rightarrow \lambda_{\pm}^{-1}E_{\pm}$.  We therefore get an
infinite tower of states $E^{(n)}_{+/-}$ obeying the geometric progression
\begin{equation}
E^{(n)}_{+}=\epsilon_+ \lambda_{+}^{-n}, \; \; E^{(n)}_{-}=\epsilon_- \lambda_{-}^{-n},
\label{geometric}
\end{equation}
for some negative energy constants $\epsilon_{+/-}$. 
We note that the case $\alpha = 2$ corresponds to a Hamiltonian of
the form
\begin{equation}
H(p,r)=-\pi J_0|p|+
 \frac{V_0}{|r|},
\end{equation}
which, for $J_0<0$ and $V_0<0$, is analogous to a relativistic fermion with
attractive Coulomb interactions.  This system is therefore directly related
to the scale anomaly recently proposed in graphene for Dirac fermions and
attractive Coulomb interactions \cite{Ovdat:2017lho}. 

For the cases where $\delta_+$ and $\delta_-$ are not real, the wave functions
at zero energy are
\begin{align}
\psi_{\rm {+}}(r)  = |r|^{-{\rm Im}\,\delta_+}\cos[{\rm Re}\,\delta_{+} \ln(|r|)],
\;  \; \;\psi_{\rm {-}}(r)
= |r|^{-{\rm Im}\,\delta_-}\sgn(r)\cos[{\rm Re}\,\delta_{-} \ln(|r|)].
\end{align} 
Under the scale transformations $r \rightarrow \lambda_{\pm} r$, the wave
functions scale homogeneously if
we let\begin{align}
\lambda_{+} = \exp (\pi/{\rm Re}\,\delta_{+}), \; \;
\lambda_{-} = \exp (\pi/{\rm Re}\,\delta_{-}).
\end{align}
 
\subsection*{Time fractals}
For the purposes of this discussion, we consider
the immobile boson localized at $r=0$ as a static source and consider only
the wave function of the second boson, which can occupy sites $r \ne 0$.
We start with an initial state 
\begin{equation}
|S\rangle = \sum_{n=0}^{N-1} |\psi^{(n)}_+\rangle,
\end{equation}
where we sum over the lowest $N$ even-parity bound states $|\psi^{(n)}_+\rangle$
 with
equal weight.  
The phase convention for each $|\psi^{(n)}_+\rangle$
is chosen
so that the tail of the wave function is real and positive at large $r$.
This state can be decomposed into position eigenstates
\begin{equation}
| S \rangle = \sum_{r \ne 0} S(r)|r\rangle.
\end{equation}
 
On a classical computer we can produce time fractals by computing the amplitude
\begin{align}
A(t) &= {\rm Re} [Z(t) ],
\end{align}
where
\begin{align}
Z(t) &= \langle S | \exp(-iHt) | S \rangle.
\end{align}
The experimental realization of time fractals on a trapped ion quantum system
requires more effort.  In order to compute the time evolution of the state
$| S \rangle$, we define
a product of single-qubit rotations 
\begin{equation}
U(\epsilon) = \prod_{r\ne 0} \exp[-i\epsilon\sigma^y_r S(r)],
\end{equation}
for some infinitesmal real parameter $\epsilon$.  With the immobile boson
still fixed at
$r=0$, let us denote the normalized state with no mobile bosons at all as
$| o \rangle$.
The action of $U$ on
the state $| o \rangle$ produces a wave function with an indefinite number
of mobile bosons.  We find that
\begin{equation}
U(\epsilon)| o \rangle = \left[1 - \frac{\epsilon^2}{2} \langle S|S\rangle
+ O(\epsilon^3)\right] |o\rangle +\epsilon
|S\rangle + O(\epsilon^2 )|X\rangle,
\end{equation}As mentioned in the main text, we have subtracted a constant
from the Hamiltonian  so that the energy of $|o\rangle$  is  zero.  Hence,
   
\begin{equation}
\langle o |\exp[-iHt] |o\rangle=\langle o |o\rangle=1.
\end{equation}
We now measure
\begin{equation}
B(\epsilon,t) = |\langle o | U^\dagger(\epsilon) \exp[-iHt] U(\epsilon)|
o \rangle|^2,
\end{equation}
This can be viewed as a quantum measurement of the projection operator $|o
\rangle \langle o |$ on the state 
\begin{equation}
U^\dagger(\epsilon) \exp[-iHt] U(\epsilon)|
o \rangle.
\end{equation}
If we deconstruct $B(\epsilon,t)$ into powers of $\epsilon,$ we get
\begin{equation}
B(\epsilon,t) =\left|1-\epsilon^2\langle S|S\rangle +\epsilon^2 Z(t)+O(\epsilon^3)\right|^2.
\end{equation}
We then obtain
\begin{align}
B(\epsilon,t) & =1-2\epsilon^2\langle S|S\rangle+\epsilon^2[Z(t)+Z^{*}(t)]+O(\epsilon^3)
\nonumber \\
 &=1-2\epsilon^2\langle S|S\rangle+2\epsilon^2  A(t)+O(\epsilon^3), 
\end{align}
and we can thus determine the desired amplitude $A(t)$.

To our knowledge this is the first instance of the concept of time fractals
appearing in the literature.  However a recent preprint \cite{C_Gao:2019}
appeared a few weeks after our preprint was posted discussing a similar concept
which they called dynamical fractals. 

\bibliography{References}

\end{document}